\documentclass[12pt]{iopart}
\def\be{\begin{equation}}
\def\ee{\end{equation}}

\def\ra{\rangle}

\def\o#1{\overline{#1}}
\def\rN{\rm N}

\def\ZZ {{\sf Z\!\!Z}}

\usepackage{iopams}  
\begin{document}
% Journal identifier can be put here if required, e.g.
\jl{6}

\begin{flushright}
  hep-th/9908155\\
  HUB-EP-99/46\\
  CERN-TH/99-257\\
\end{flushright}
\vskip 0.3cm

\title[Aspects of Type 0 String Theory]{Aspects of Type 0 String Theory
\footnote[3]{Talk presented by R.B. at Strings '99, Potsdam,
July 19-24 1999.}}

\author{R. Blumenhagen\dag\ A.Font\ddag\ A. Kumar$\sharp$\ 
and D. L\"ust\dag}

\address{\dag\ Humboldt-Universit\"at zu Berlin, Institut f\"ur 
Physik,Invalidenstrasse 110, 10115 Berlin, Germany.}

\address{\ddag\ Departamento de F\'{\i}sica, Facultad de Ciencias,
 Universidad Central de Venezuela, A.P. 20513, Caracas 1020-A, Venezuela.} 

\address{$\sharp$\ Theory Divison, CERN, CH-1211, Geneva 23, Switzerland;
 Institute of Physics, Bhubaneswar 751 005, India}

\begin{abstract}
A construction of compact tachyon-free orientifolds of 
the non-supersymmetric
Type 0B string theory is presented. Moreover, we study effective 
non-supersymmetric 
gauge theories arising on self-dual D3-branes in Type 0B orbifolds
and orientifolds.
\end{abstract}


\submitted

\maketitle

\section{Introduction}
\setcounter{equation}{0} 

If one would like superstring theory to give rise to  some 
testable predictions
in the low energy regime, at some point one has to explain how supersymmetry
is broken somewhere  between the Planck and the weak scale. 
In another approach one might contemplate to start already with a 
non-supersymmetric string theory at the Planck scale. 
However, generically non-supersymmetric string theories are plagued 
with problems luckily absent in supersymmetric string theories. 
As in the bosonic string theory, in most non-supersymmetric theories
tachyons appear, which in the best of all imaginable scenarios indicate 
a phase 
transition into some stable background.  Moreover, without supersymmetry
a big cosmological constant is generated by loop corrections in 
conflict with the small value of the cosmological constant we observe
at least in our universe. Non-supersymmetric string theories with
vanishing cosmological constant have been studied in \cite{KKS}.
Moreover, the cosmological constant serves as a dilaton potential leading
to the request for stabilization of the dilaton, i.e.
of the string coupling constant.
Analogously,  
scalar potentials for all other moduli are usually generated, which 
must be stabilized, as well. Concerning establishing string dualities,
one is also on less solid ground  as compared to 
supersymmetric string theories.
In particular, there is no notion of BPS like objects and masses 
receive quantum corrections. 

In this paper we will mainly deal with the first of the problems
mentioned above, namely we will construct non-supersymmetric
models free of tachyonic modes.
More concretely we will consider special orientifolds of the Type 0B
string theory, where the tachyon does not survive the orientifold projection.
We will discuss models in ten, six and four space-time dimensions. 
In all these models the second of the problems from above is present, as a
dilaton tadpole is generated on the disc world-sheet.   
In the second part we will study the effective gauge theories arising
on self-dual D3-branes in non-compact Type 0B orbifolds and
orientifolds. This approach was initiated in \cite{KLEB1,KLEB2} where 
a generalization of AdS-CFT duality to the the supersymmetric
Type 0B backgrounds was presented. Here we will mainly focus on 
the CFT side.  
\smallskip

This talk is based  on \cite{BFL1,BFL2,BK} where more details 
can be found.

\section{Non-tachyonic orientifolds of Type 0B}

There exist two different constructions of  Type 0B string theory. 
It can be
regarded first as the superstring with the projection
\be
\label{GSO}
P={1\over 2}(1+(-1)^{F_L+F_R})\,,
\ee
and second as Type IIB divided out by the space-time fermion number
$(-1)^{F_S}$. This leads to a modular invariant partition function containing
a tachyon, a graviton,
a dilaton and an antisymmetric two form in the NSNS sector and
two further scalars, two antisymmetric 2-forms 
and a 4-form in the RR sector. Except the tachyon all these states
are massless.
Since compared to the Type IIB models all fields in the RR sector are 
doubled, the D-brane content is doubled, as well. The explicit 
form of the corresponding boundary states was derived in \cite{BG1}.
The two types of D-branes of the same dimension $p$ are denoted as
D$p$- and D$p'$-branes in the following. 
Studying the annulus amplitude in tree channel using the boundary states
and transforming back into loop channel one 
derives that for open strings stretched between 
the same type of D$p$-branes the world-sheet fermions have half-integer
mode expansion, whereas for open strings stretched between a 
D$p$- and a D$p'$ -rane they have integer mode expansion. 
In the first case one gets space-time bosons and the second 
case  space-time fermions. Thus, even though the closed string Type 0B
theory is purely bosonic, fermions appear  in the D-brane sector.
 
As was first realized in \cite{SAGBI,SAGN} there exist different
orientifold projections in Type 0B. The usual orientifold by the
world-sheet parity transformation $\Omega$ yields a model
still containing the tachyon. 
In \cite{BG1,SAGBI} it was shown that the dilaton tadpole can be cancelled
by introducing 32 D9 and 32 anti D9-branes into the background leading
to a gauge group SO(32)$\times$SO(32) and an open string tachyon in
the bi-fundamental representation of the gauge group. 
The introduction of anti-branes was necessary to
cancel the dangerous RR tadpole. The model above was conjectured to be 
strong-weak dual to the bosonic string compactified on the
root lattice of SO(32). Moreover, the orientifold model still contains
a tachyonic tadpole, which can only be cancelled by introducing instead
16 D9, 16 D9', 16 $\o{\rm D9}$ and 16 $\o{\rm D9}$'-branes in the 
background. This model has gauge group SO(16)$\times$SO(16)$\times$
SO(16)$\times$SO(16) and was conjectured to be related to the 
tachyon-free SO(16)$\times$SO(16) heterotic string in \cite{BK}.

An independent model is defined by combining the world-sheet parity
transformation with the right moving word-sheet fermion number operator
$\Omega'=\Omega (-1)^{f_R}$. In this case there appear RR tadpoles
in the Klein bottle amplitude which can be cancelled by 32 D9- and
32 D9'-branes. However,  from the annulus amplitude  there remains
an uncancelled dilaton tadpole which leads to a shift in the 
expectation values of the background fields via the Fischler-Susskind
mechanism \cite{FS}.
The massless closed string spectrum is given by the bosonic part 
of Type IIB spectrum and the self-dual D9-branes support a U(32) gauge
group with Majorana-Weyl fermions in the ${\bf 496}\oplus\o{\bf 496}$ 
representation. The closed and open string together cancel the
$R^6$ and $F^6$ anomalies. Note, that in contrast to Type I string theory
this orientifold still contains Dp-branes for every odd number of p.
This orientifold model was conjectured to be related to the tachyonic 
U(16) heterotic string theory.
For completeness we mention that there exists a third orientifold projection.
The world-sheet parity
transformation is combined with the right moving space-time  fermion number 
operator leading to a tachyon tadpole in the Klein bottle amplitude.

The question is whether the absence of tachyons holds under
toroidal orbifold compactifications.  
It was argued in \cite{ANG} that in  general this it not the case, as
the world-sheet parity transformation exchanges a $g$ twisted sector with
the $g^{-1}$ twisted sector implying that some linear combinations
of the twisted sector tachyons survive the projection. Orientifolds
in which the twisted sectors are invariant under $\Omega$ were discussed
in \cite{KST,BGK} and might have  interesting non-supersymmetric 
generalizations.

From the set of supersymmetric orbifolds there are however two models
in which all tachyons are projected out. In six flat space-time dimensions
one has the $\ZZ_2$ orbifold of $T^4$ with action $z_i\to -z_i$ for each
of the two complex coordinates. Performing the tadpole cancellation
condition, one realizes that all RR tadpoles can be cancelled by introducing
32 self-dual D9-branes and 32 self-dual D5-branes in the background.
At the massless level the closed string sector contains the graviton, 
the dilaton, 20 self dual
2-forms, 4 anti-self dual 2-forms and 98 further scalars. 
In the open string sector one gets gauge group
G=U(16)$\times$ U(16)$\vert_9\times$ U(16)$\times$ U(16)$\vert_5$, 
where the first two factors live on the D9-branes and the second two on 
the D5-branes. The open strings connecting the various kinds of D-branes and
D'-branes yield the following massless matter states
\smallskip
\be
\begin{array}{lclr}
4\times\{({\bf 16},{\bf\o{16}};
    {\bf 1},{\bf 1}) + 
   ({\bf\o{16}}, {\bf{16}};{\bf 1},{\bf 1}) + 
   ({\bf 1},{\bf 1};{\bf 16},{\bf\o{16}}) + 
 ({\bf 1},{\bf 1};{\bf\o{16}},{\bf 16}) \}_{1,1} \\
2\times\{({\bf 16},{\bf 1};{\bf\o{16}},{\bf 1}) + 
({\bf\o{16}},{\bf 1};{\bf {16}},{\bf 1}) +
 ({\bf 1},{\bf 16};{\bf 1},{\bf\o{16}}) +
 ({\bf 1},{\bf\o{16}};{\bf 1},{\bf 16}) \}_{(1,1)}\\
2\times\{({\bf 120}\oplus {\bf\o{120}},{\bf 1};{\bf 1},{\bf 1}) + 
        ({\bf 1},{\bf 120}\oplus {\bf\o{120}};{\bf 1},{\bf 1}) + \\
  \quad\quad\ ({\bf 1},{\bf 1};{\bf 120}\oplus {\bf\o{120}},{\bf 1})+
    ({\bf 1},{\bf 1};{\bf 1},{\bf 120}\oplus {\bf\o{120}}) \}_{(1,2)} \\
2\times\{({\bf 16},{\bf 16};{\bf 1},{\bf 1}) + 
({\bf\o{16}},{\bf\o{16}};{\bf 1},{\bf 1}) + 
 ({\bf 1},{\bf 1};{\bf 16},{\bf 16}) + 
({\bf 1},{\bf 1};{\bf\o{16}},{\bf\o{16}}) \}_{(2,1)}\\
1\times\{({\bf 16},{\bf 1};{\bf 1},{\bf 16}) + 
         ({\bf\o{16}},{\bf 1};{\bf 1},{\bf\o{16}}) +
            ({\bf 1},{\bf 16};{\bf 16},{\bf 1}) + 
        ({\bf 1},{\bf\o{16}};{\bf\o{16}},{\bf 1}) \}_{(1,2)}
\end{array}
\ee
\smallskip
where the index indicates the representation under the SU(2)$\times$SU(2)
Lorentz-group in six-dimensions. 
For the complete closed and open spectrum both the $R^4$ and the 
$F^4$ anomaly cancels.

In four flat space-time dimensions the orientifold on $T^6/\ZZ_3$ 
is free of tachyons. Note, that it is only for the $\ZZ_3$ orbifold 
that the twisted sector ground state energy vanishes. 
In this case all RR tadpoles can be cancelled by self-dual D9-branes.
In the closed string sector one gets the graviton plus 10 scalars,
including the dilaton and internal metric moduli, and additional 
20 RR-scalars and one vector that arises from the 4-form. In the twisted
sector appears another 27 NS-NS massless scalars and 54 R-R massless scalars.
In the open string sector one obtains gauge group 
G=U(12)$\times$ U(12)$\times$ U(8) with bosonic and fermionic matter
\smallskip
\be
\begin{array}{lclr}
3\times\{ ({\bf 12}, {\bf \o{12}}, {\bf 1}) + 
({\bf 1}, {\bf 12}, {\bf \o{8}}) +
({\bf \o{12}}, {\bf 1}, {\bf 8}) + c.c.\}_{B} \\
1\times\{ ({\bf 12}, {\bf 12}, {\bf 1}) + ({\bf 1}, 
{\bf 1}, {\bf 28}) + ({\bf\o{12}}, {\bf\o{12}}, {\bf 1} ) + 
({\bf 1}, {\bf 1}, {\bf\o{28}}) \}_{L} \\
3 \times \{ ({\bf 66},{\bf 1},{\bf 1}) + ({\bf 1},{\bf 12},{\bf 8}) + 
({\bf 1}, {\bf\o{66}}, {\bf 1}) + 
({\bf\o{12}}, {\bf 1}, {\bf\o{8}}) \}_{L}.
\end{array}
\ee
\smallskip
This spectrum is chiral and free of non-Abelian gauge anomalies. 
Concerning the $U(1)$ factors, there is one
non-anomalous and two anomalous combinations whose anomaly could presumably
be cancelled by a generalized Green-Schwarz mechanism.

So far we have only considered orbifolds preserving some supersymmetry
in the Type IIB setting. Since Type 0B is non-supersymmetric anyway, we
are free to consider more general orbifold actions. One non-tachyonic
four dimensional examples of this sort was discussed in \cite{BK}. 
One simply takes $T^6$ and divides out by the $\ZZ_2$ action $z_i\to z_i$
for all three complex coordinates. Note, that in Type IIB this
orbifold would not satisfy level-matching. In particular, the
level-matching condition would be violated in the NS-R sector, but
precisely this sector is absent in Type 0B. One subtlety arises in the 
Ramond sector where the action of the $\ZZ_2$ on the ground states
is 
\be
R|s_1\, s_2\, s_3\, s_4\ra=e^{\pi i(s_2+s_3+s_4)}
                  |s_1\, s_2\, s_3\, s_4\ra =\pm i|s_1\, s_2\, s_3\, s_4\ra
\ee
with $s_i=\pm 1/2$.
This action is rather $\ZZ_4$ than $\ZZ_2$. In the closed string sector
the left-moving Ramond sector is always paired with the right-moving
Ramond sector, so that the action is really $\ZZ_2$, but in the open 
string sector
this $\ZZ_4$ action on the Ramond ground  states has to compensated by
a $\ZZ_4$ action on the Chan-Paton factors. Moreover, the Klein bottle
amplitude for this model leads to tadpoles which can be cancelled by 32
self-dual D9 -and self-dual D3-branes. The technical aspects of this
model are discussed in length in \cite{BK} and everything works out
just right to eventually lead to a chiral but anomaly free massless 
spectrum. The closed string sector contributes
the graviton, the dilaton and 117 further scalars and the branes
support a gauge group G=U(16)$\times$ U(16)$\vert_{9}\times$ U(16)
$\times$  U(16)$\vert_3$ with matter
\smallskip
\be
\begin{array}{lclr}
6\times\{ ({\bf 16},{\bf\o{16}};
    {\bf 1},{\bf 1}) + 
   ({\bf\o{16}}, {\bf{16}};{\bf 1},{\bf 1}) + 
   ({\bf 1},{\bf 1};{\bf 16},{\bf\o{16}}) + 
 ({\bf 1},{\bf 1};{\bf\o{16}},{\bf 16}) \}_B \\
4\times\{ ({\bf 120} ,{\bf 1};{\bf 1},{\bf 1}) + 
        ({\bf 1},{\bf 120};{\bf 1},{\bf 1}) + 
  ({\bf 1},{\bf 1};{\bf 120},{\bf 1})+
    ({\bf 1},{\bf 1};{\bf 1},{\bf 120}) \}_L \\
4\times\{ ({\bf\o{16}},{\bf\o{16}};{\bf 1},{\bf 1}) + 
     ({\bf 1},{\bf 1};{\bf\o{16}},{\bf\o{16}}) \}_L \\
1\times\{ ({\bf 16},{\bf 1};{\bf 16},1) + 
               ({\bf 1},{\bf 16};{\bf 1},{\bf 16}) \}_L 
\end{array}
\ee
\smallskip
Thus, we have constructed a couple of tachyon-free non-supersymmetric
orientifolds in various space-time dimensions. The spectra we obtained 
are of course only correct at tree-level. Quantum corrections will
lead to an effective potential for the brane moduli, the minima of which 
will determine the final positions of the branes and therefore
the gauge symmetry. In the computation of the massless spectra we have 
implicitly assumed
that all branes of the same dimension  lie on top of each other. Taking
quantum corrections into account this might simply not be a minimum
of the effective potential. The analysis of \cite{BD} might suggest
that maybe all non-supersymmetric models are driven by quantum induced
potentials for the moduli  to supersymmetric configurations, 
so that maybe all
stable vacua of string theory are supersymmetric. However, in particular
for the non-supersymmetric orientifold $T^6/\ZZ_2$ we find it hard to imagine
a candidate supersymmetric vacuum to which it might flow. 
 
\section{Effective gauge theories on D3-branes}

It was suggested in \cite{POLY} to study the dynamics of non-supersymmetric
gauge theories by viewing them as effective theories arising  on 
D3-branes in Type 0B string theory. 
This idea was worked out in \cite{KLEB1}, where the low energy effective 
action of Type 0B was computed. Since this action contains a coupling
of the tachyon to the square of the RR 5-form field strength, introducing
RR 5-form flux into the background can cure the tachyonic instability. 
For self-dual
D3-branes the tachyon decouples and a generalized AdS-CFT correspondence
was established in \cite{KLEB2} which turned out to be stable as long
as the 't Hooft coupling satisfies $\lambda=g_{YM}^2 {\rm N}<100$.
 
The effective theory arising on N parallel self-dual D3-branes in
Type 0B has gauge group SU(N)$\times$SU(N) and three complex bosons
in the $({\bf Adj},{\bf 1})$ and $({\bf 1},{\bf Adj})$ 
representation. Moreover, there arise four Weyl-fermions in the
$({\bf N},{\bf\o{N}})+({\bf\o{N}},{\bf N})$ representation of the 
gauge group. 
It is easily shown that the one-loop $\beta$-function vanishes identically
for this matter spectrum and the two-loop $\beta$-function vanishes only
in the large N limit. 
Thus, as expected for non-supersymmetric gauge theories the AdS-CFT
correspondence only tells us that the gauge theory is conformal only in 
the large N limit. There are non-zero 1/N corrections which correspond
to string loop corrections to the AdS$_5$ geometry. 
The exact vanishing of the
one-loop $\beta$-function is related to the vanishing of the annulus 
amplitude for self-dual D-branes in Type 0B. The non-vanishing of the 
two-loop $\beta$-function  therefore tells us, that the two loop string
partition function will be non-zero.
In fact, the observation that Type 0B can be regarded as an orbifold of 
Type IIB and  the general arguments given in \cite{BKV} imply that 
all higher loop $\beta$-functions vanish in the large N limit. 
Unfortunately, it is so far out of reach of our methods to decide 
whether there exists
a non-trivial fixed point of the entire renormalization group flow even
for finite value of N. 
In the following we will study what kinds of gauge theories one gets 
by taking orbifolds and orientifolds of the original model. 

\subsection{Orbifolds}

We are placing N self-dual D3-branes on a non-compact $\ZZ_K$ orbifold.
Similar to the compact case discussed in
Section 2 we are free to choose orbifolds preserving {\cal N}=2 \cite{BCR}, 
{\cal N}=1 or even no supersymmetry at all. 
We compute the annulus amplitude 
\be
A= \int_0^\infty \, {{dt}\over t} \, 
                 {\rm Tr}_{open} \left[ \left( {1\over K}\sum_{i=0}^K 
          \Theta^i\right) e^{-2\pi t L_0} \right]
\ee
for self-dual D3-branes and require the absence of both linear and
logarithmic ultraviolet divergences. This leads to conditions for
the action $\gamma_{\Theta^i}$  of the symmetry on the Chan-Paton factors.
Since we are in a non-compact setting and have the same number of D3- and
D3'-branes, there arises no massless or tachyonic tadpole in the untwisted
sector.
Therefore there exists no restriction on $\gamma_{1}$ and the number of 
D3-branes is  arbitrary. In the orbifold case, the twisted tadpole conditions
are simply $\gamma_{\Theta^i}=0$ for $i=1,\ldots,K$. Explicit
results can be found in \cite{BFL2}, here we just mention the general 
features. Since the annulus amplitude vanishes exactly, all spectra
are bose-fermi degenerated and the one-loop
$\beta$-functions are zero, as well.
At two loop order one encounters 1/N corrections. Moreover, all gauge groups
are of the type $\prod_i {\rm SU}(n_i)^2$, thus every unitary gauge group 
appears twice. In order to get
more general gauge groups, like single SU(n) factors or orthogonal and
symplectic gauge groups, one has to  consider orientifolds.  
 
\subsection{Orientifolds}

As in the compact case, one has to distinguish between two different 
orientifold projections. One can consider first non-compact orientifolds 
with the
T-dual of the $\Omega$ operation and second orientifolds 
with the T-dual of $\Omega (-1)^{f_R}$. 
Note, that we need to take the T-dual operators, 
$\omega=\Omega J (-1)^{F_L}$ and
$\omega'=\Omega (-1)^{f_R} J (-1)^{F_L}$ with $J:z_{1,2,3}\to -z_{1,2,3}$,
as we are interested in effective
four dimensional models and therefore need D3-branes instead of D9-branes.

\subsubsection{$\omega$ orientifolds}
These are orientifolds by the group $G+\omega G$ with 
$G=\ZZ_K$. In the compact case the untwisted RR tadpole cancellation 
condition forced us to introduce anti-branes and therefore open
string tachyons into the background. However in the non-compact case
there are only twisted sector RR tadpoles, which 
are cancelled by choosing $\gamma_{\Theta^i}=0$ for $i=1,\ldots,K$.
Therefore, there is no need to introduce anti-branes and one gets 
stable gauge theories on the self-dual D3-branes. 
Here we would like to only discuss the easiest case with trivial G, more
complicated examples can be found in \cite{BFL2}. 
For $G=1$ we find the two allowed gauge groups
G=SO(N)$\times$ SO(N) and G'=SP(N)$\times$ SP(N)
and three complex bosons
in the $({\bf Adj},{\bf 1})$ and $({\bf 1},{\bf Adj})$ 
representation. Moreover, we have  four Weyl-fermions in the
$({\bf N},{\bf N})$  representation of the gauge group. 
In all orientifold  models the one-loop $\beta$-function of the
gauge coupling vanishes only
in the large N limit, namely in this case it is $b_1=0\, \rN\mp16/3$.
From the string theoretic point of view this is simply due to the fact
that, even though the annulus amplitude is still vanishing, the 
M\"obius amplitude
is non-zero. Thus, there is  a one-loop cosmological constant generated,
which causes the dilaton and therefore the gauge coupling to run.
Since the M\"obius amplitude is 1/N suppressed against
the annulus amplitude, this running is a 1/N effect in the 
one-loop $\beta$-function. For the two-loop $\beta$-function we obtain
$b_2=\pm {64/3} ( \rN\mp {1/2} )$ where the N$^2$ term vanishes. 
Note, that $b_1$ and $b_2$ have opposite sign so that there is a chance
to find a non-trivial fixed point at some finite value of g.  Of course,
we can not prove that such a fixed point really exists.
Going to more general  orbifold groups $G$
does not change the general patterns mentioned above, 
it only produces more general gauge and matter contents. 

\subsubsection{$\omega'$ orientifolds}
These are orientifolds by the group $G+\omega'\, G$ with 
$G=\ZZ_K$. One finds the same RR-tadpole cancellation conditions
as in the corresponding Type IIB orientifold
\be
\label{tad}
 {\rm Tr}(\gamma_{\Theta^{2k}})=\pm {1\over \prod_{i=1}^3\,
           {\cos}\left( {\pi kv_i\over K}\right) },
\ee
where $\Theta$ denotes the generator of $\ZZ_K$ and acts on the three complex
coordinates transversal  the D3-branes as
$\Theta\, z_i\to {\rm exp}(2\pi i v_i/K) z_i$.
There arises  no untwisted dilaton tadpole but there are new twisted
sector NSNS tadpoles. 
Since the trace of $\gamma_{1}$  is proportional to N and the relation
(\ref{tad}) holds,  they are 1/N suppressed.  
Choosing the trivial orbifold group G leads to a gauge group
G=SU(N), three complex bosons in the $({\bf Adj})$  and  
four Weyl-fermions in the $({\bf A}+{\bf \o{A}})$ or
$({\bf S}+{\bf \o{S}})$  representation of the 
gauge group. For the one-loop $\beta$-function we find $b_1=0\, \rN\pm 16/3$
and the two-loop $\beta$-function consistently vanishes in the large 
N limit. For this class of orientifolds generally one finds an odd
number of SU(n) gauge factors but no orthogonal or symplectic gauge
groups. 

\section{Conclusions}

In \cite{BFL2,AK} the T-dual description of the models presented
here was discussed. These are the Type 0A generalizations of
Hanany-Witten set-ups including self-dual D4-branes stretched between
NS5-branes. The rules for determining the massless spectra are
easily derived and one finds as the general feature,  that the 
one-loop $\beta$-function of the coupling in
the non-supersymmetric gauge theory is the same as the one-loop 
$\beta$-function 
in the corresponding supersymmetric Type IIA gauge theory.
It is this property which distinguishes the Type 0 models from the
general class of non-supersymmetric models discussed in \cite{KS}. 
It would be interesting to see how far the methods and results derived
in the supersymmetric case could  be generalized to the non-supersymmetric 
case at least qualitatively. In particular one would like to embed
the Hanany-Witten set-ups into M-theory by using the conjectured
duality \cite{BG2} of Type 0A to M-theory on $S^1/(-1)^{F_S} S$ 
where $S$ denotes the shift around the half-circle. 

Moreover, a better understanding of the process of tachyon condensation
and which models are related by that is desirable \cite{KKS2}. 
It was proposed  in \cite{FV} to solve the gauge hierarchy problem via 
conformal field theories. The realization of this idea
strongly depends on the
existence of non-trivial renormalization group fixed points in 
non-supersymmetric gauge theories. 
It would be nice if string theoretic methods could give some
new insights into the existence of such fixed points.

\end{document}